\documentclass[
superscriptaddress,
twocolumn,
showpacs,
bibnotes,
amsmath,
amssymb,
aps,
longbibliography,
prl,
floatfix,
]{revtex4-2}
\usepackage{graphicx} 

\usepackage{dcolumn}
\usepackage{float}
\usepackage{bm}
\usepackage{cleveref}
\usepackage{tabularx}
\usepackage{booktabs}
\usepackage{silence}
\usepackage{ulem}
\usepackage{color}
\usepackage{soul}
\usepackage{xcolor}
\usepackage{hyperref}

\newcommand{\Fal}[1]{Fe$_4$Al$_{13}$}

\begin{document}

\title{Dilute Paramagnetism and Non-Trivial Topology in Quasicrystal Approximant Fe$_4$Al$_{13}$} 
\author{Keenan E. Avers}
    \email[]{kavers@umd.edu}
    \affiliation{Maryland Quantum Materials Center and Department of Physics, University of Maryland, College Park, Maryland, USA}

\author{Jarryd A. Horn}
    \affiliation{Maryland Quantum Materials Center and Department of Physics, University of Maryland, College Park, Maryland, USA}
   
\author{Ram Kumar}
    \affiliation{Maryland Quantum Materials Center and Department of Physics, University of Maryland, College Park, Maryland, USA}
 
\author{Shanta R. Saha}
    \affiliation{Maryland Quantum Materials Center and Department of Physics, University of Maryland, College Park, Maryland, USA}

\author{Yuanfeng Xu}
    \affiliation{Department of Physics, Princeton University, Princeton, NJ 08544, USA}
    \affiliation{Center for Correlated Matter and 
 School of Physics, Zhejiang University, Hangzhou 310058, China }

\author{B. Andrei Bernevig}
    \affiliation{Department of Physics, Princeton University, Princeton, NJ 08544, USA}
    
\author{Peter Zavalij}
\affiliation{Department of Chemistry, University of Maryland, College Park, Maryland 20742, USA}

\author{Johnpierre Paglione}
    \affiliation{Maryland Quantum Materials Center and Department of Physics, University of Maryland, College Park, Maryland, USA}
    \affiliation{Canadian Institute for Advanced Research, Toronto, Ontario M5G 1Z8, Canada}
    \email{paglione@umd.edu}
    
\date{\today}

\begin{abstract}
 A very fundamental property of both weakly and strongly interacting materials is the nature of its magnetic response. In this work we detail the growth of crystals of the quasicrystal approximant \Fal{} with an Al flux solvent method. We characterize our samples using electrical transport and heat capacity, yielding results consistent with a simple non-magnetic metal. However, magnetization measurements portray an extremely unusual response for a dilute paramagnet and do not exhibit the characteristic Curie-Weiss behavior expected for a weakly interacting material at high temperature. Electronic structure calculations confirm metallic behavior, but also indicate that each isolated band near the Fermi energy hosts non-trivial topologies including strong, weak and nodal components, with resultant topological surface states distinguishable from bulk states on the (001) surface. With half-filled flat bands apparent in the calculation but absence of long-range magnetic order, the unusual paramagnetic response suggests the dilute paramagnetic behavior in this quasicrystal approximant is surprising and may serve as a test of the fundamental assumptions that are taken for granted for the magnetic response of weakly interacting systems.
 \end{abstract}

\maketitle





\section{Introduction}

The magnetic response of a material is among one of the most fundamental properties used to understand the physics of a many-body system. Many strongly interacting materials have exotic magnetic states and behaviors at low temperatures, such as frustrated antiferromagnetism in triangular lattices \cite{Khatua2022PRB, Avers2021PRB}, quantum criticality in heavy Fermion systems \cite{Ajesh2022PhysRevB.106.L041116, Qimiao2010Science}, and Skrymions in broken inversion symmetric lattices \cite{Munzer2010PhysRevB.81.041203FeCoSiSkyrmion}. In contrast, especially for weakly correlated/interacting systems at sufficiently high temperatures, all these systems should end up as a paramagnet and obey the Curie-Weiss law \cite{Mugiraneza2022ComPhys}.

In this work we draw attention to our crystals of Al flux-grown \Fal{} and its characterization as an example of a weakly interacting system. The material exhibits very simple electrical transport consistent with a non-magnetic metal and heat capacity contributions of only conduction electrons and lattice phonons, with no indication of any electronic or structural phase transitions. The material does, however, have a complicated monoclinic crystal structure with a large unit cell that hosts off-stoichiometric paramagnetic Fe impurities. Although the Curie-Weiss law should be valid for such a dilute paramagnet, to our surprise the magnetic response of the material exhibits a complicated temperature and field dependence. Our results conflict with  the basic assumptions of a weakly interacting dilute paramagnet.  

As Fe and Al are extremely common and inexpensive elements, it is no surprise that Fe-Al binaries have been extensively investigated \cite{Liu:pj4005, Grin1994Kristal, DING2022114726, Ellner:js0001, ZIENERT2017848, KazukiTobita2016MF201607, MINGJIAN1988L23}. It is important to note that \Fal{} borders on formation of a quasicrystal \cite{Agrosi2024QuasicrystalMeteor}: a crystal lattice with a nominally forbidden long-range arrangement of atoms that exhibits no discrete translational symmetries, and lacking an underlying unique unit cell \cite{Shechtman1984PhysRevLett.53.1951}. However, quasicrsytals still maintain a global sense of discrete rotational symmetries \cite{He2016QuasicrystalDeca, Mihal2020PhysRevResearch.2.013196}, in contrast to fully disordered glassy solids \cite{SUN20081159}. A fundamental assumption in condensed matter physics is the notion that a material exhibits rotational and translational symmetries of a crystalline structure. These  in turn allow a symmetry-based starting point for understanding electrical and magnetic properties from a microscopic level \cite{dresselhaus2007group}. It is an open question as to what electronic and magnetic \cite{LABIB2024101321, Nawa2023PhysRevMaterials.7.054412, Deguchi2012QuasicrystalQuantumCrtic} physics are permissible when such fundamental assumptions are relaxed.
While the \Fal{} crystal structure is the subject of interest for aforementioned reasons, here we focus on unreported properties of \Fal{} as a result of using this system for a simple demonstration of flux crystal growth \cite{FQM_bible, Canfield1992GrowthOS}. 

\section{Materials and Methods}

\begin{figure*}[]
    \centering
   \includegraphics[scale=0.45]{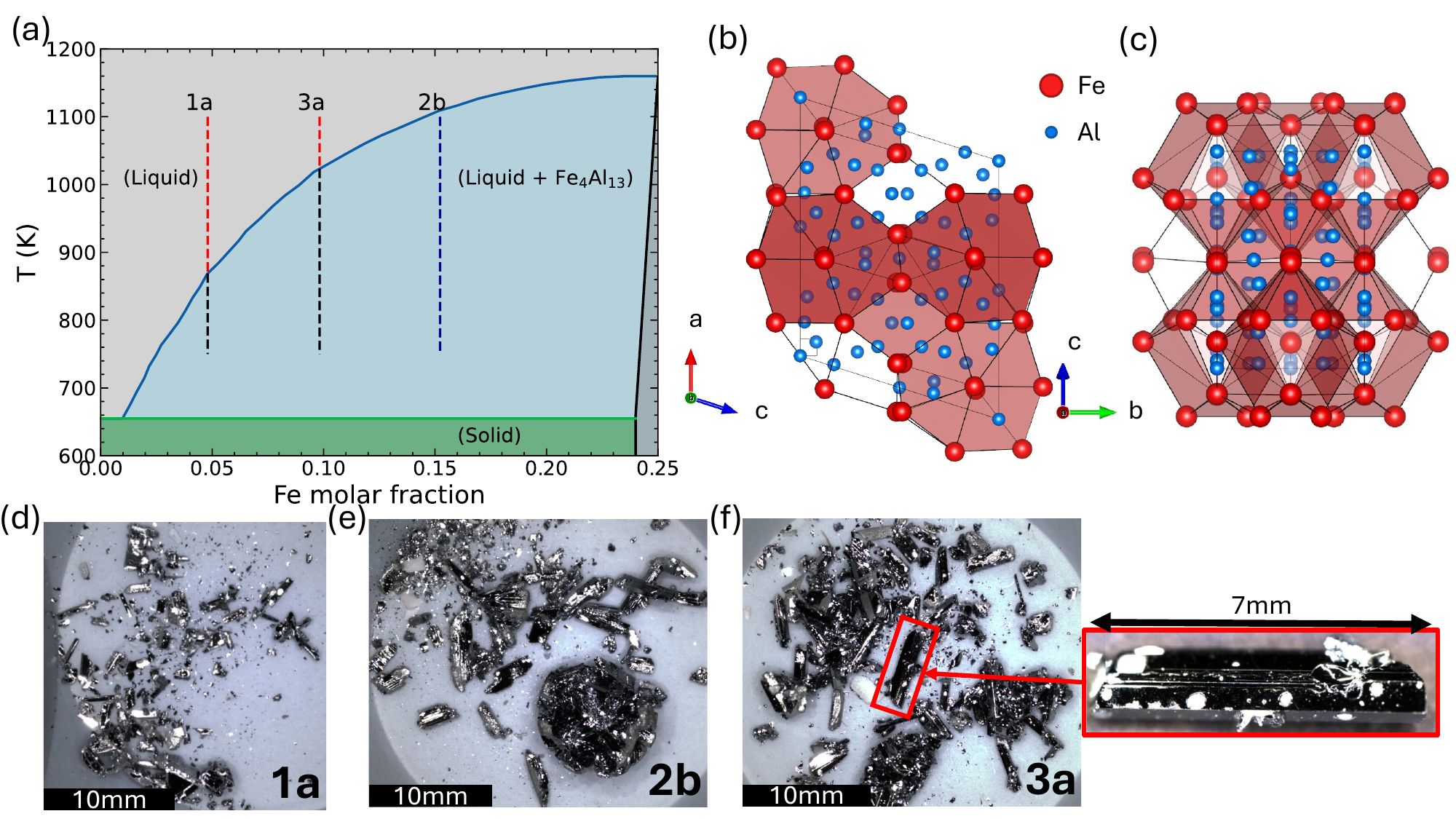}
   \caption{Systematic study of molten Al flux growth of Fe$_{4}$Al$_{13}$ crystals produced by students of the 2024 Fundamentals of Quantum Materials winter school. (a) Alloy phase diagram representation of three synthesis batches with approximately 1:20, 1:10 and 1:5 Fe:Al molar ratio for batches 1a, 3a and 2b, respectively. (b-c) Unit cell of Fe$_4$Al$_{13}$ from single crystal x-ray refinement projected onto the $a$-$c$ and $b$-$c$ planes. (d-f) Photographs of the resulting crystal yield from batches 1a, 2b and 3a. }
    \label{fig1}
\end{figure*}

Crystals of \Fal{} were grown as part of a practical training session for the University of Maryland's 2024 Fundamentals of Quantum Materials Winter School (FQM2024),  with basic yields and systematic approach shown in Figure~ \ref{fig1}. The synthesis of this Fe-Al compound was carried out using an Al ``self-flux technique '', which allows crystals to nucleate in an excess of solvent composed of intrinsic elements in the flux (in this case, Al). The \Fal{} composition itself was selected for training because of its wide liquidus region and quick crystal growth. This allows a pragmatic hands-on educational introduction to the flux growth technique that demonstrates the ability to vary the relative Fe-Al composition over the liquidus range and still obtain the correct stoichiometric crystal formation. The approach also allows for systematically investigating how the morphology of the yielded crystals may change with starting Fe:Al ratio.  Elemental Fe granules (Thermo Scientific 99.98 \%) and Al shot (Alfa Aesar 99.999 \%) were loaded into 2 mL alumina crucibles and flame-sealed in quartz ampoules with Ar gas. The ampoules were heated to 1100 C$^\circ$ and held at that temperature for 6 hours. The temperature was then decreased to 750 C$^\circ$ over 6 hours, and held at that temperature until removal from the furnace and  centrifugation of the excess Al. Examples of three groups with different trial stoichiometries - labeled 1a, 3a, and 2b - are shown in Figure~ \ref{fig1}(a) on a section of the Fe-Al binary alloy phase diagram. The $ac$-plane of the monoclinic crystal structure is shown in Figure~ \ref{fig1}(b) and exemplifies its nature as bordering on formation of a decagonal quasicrystal. In contrast, the $bc$-plane view of the unit cell is a bit more regular, consistent with the $b$-axis being much shorter than the $a$- and $c$-axis. Photographs of the resultant \Fal{} crystals are shown in Figure~ \ref{fig1}(d), (e) and (f). Depending on the exact molar ratio of Fe to Al, the crystals changed size and morphology, with one example crystal reaching a length of 7~mm, which exemplifies the simplicity and relative speed of growth of this material that lends itself well to educational and demonstrative purposes. A general trend was observed from this study, where a too dilute Fe to Al molar ratio (Figure~ \ref{fig1}(d)) does not allow for sufficient cooling time below the liquidus curve to yield larger crystals nor decent mass yield, while in contrast, an Fe-rich ratio can result in incomplete dissolution and formation of polycrystalline lumps (Figure~ \ref{fig1}(e)).

\begin{table*}[] 
\caption{Single-crystal X-ray refinement parameters for Fe$_4$Al$_{13}$ measured on a Bruker D8Venture w/ PhotonIII diffractometer. Integral intensity were correct for absorption using SADABS software\cite{Krause:aj5242} using multi-scan method. Structures were solved with the ShelXT \cite{Sheldrick:sc5086} program and refined with the ShelXL program \cite{Sheldrick:fa3356} using least-square minimization. All results are consistent with monoclinic space group C2/m with 6 formula units per unit cell.}

\label{tab1}
\newcolumntype{C}{>{\centering\arraybackslash}X}
\begin{tabularx}{\textwidth}{CCCC}
\toprule
T (K) & 250 & 150 & 100  \\
\midrule
$a$ (\AA)     &   15.4659(9) & 15.440(5) &  15.447(2)   \\
$b$ (\AA)     &  8.0759(5) & 8.067(2) &  8.0677(12)  \\
$c$ (\AA)     &  12.4618(7) & 12.452(4) &  12.4458(18)  \\
$\beta$ ($^\circ$) &  107.7041(9)  & 107.728(4) & 107.701(2)  \\
$V$ (\AA$^3$) &  1482.78(15)  & 1477.3 & 1477.6(4)   \\
$\rho$ (g/cm$^3$) &  3.858  & 3.872 & 3.871 \\
R$_1$ &  0.0217 & 0.0219 & 0.0218  \\
wR$_2$ &  0.0494 & 0.0504 & 0.0490 \\

\bottomrule
\end{tabularx}
\end{table*}

While the systematic relation between starting Fe:Al ratio and crystal yield and size is relatively simple to understand, interpreting how the Fe:Al ratio impacts a sample's electrical and magnetic properties is not straightforward, as regions of the Fe-Al binary phase diagram host many alloys and compounds with wide stoichiometry ranges and degrees of metastability \cite{Shang2021SciRep_FeAlAlloys}. To this end, additional crystals of \Fal{} were grown with Fe:Al ratios of 1:5, 1:10, and 1:20 using a similar recipe as the FQM2024 students, but over a longer time frame; the temperature of 1100 C$^\circ$ was held for 24 hours and the temperature was slowly decreased by 3 C$^\circ$ per hour to 800 C$^\circ$ at which the centrifuging occurred. All further results in this manuscript are from these samples. A similar trend to the size and morphology was observed with the Fe:Al = 1:20 crystals being small needles and Fe:Al =1:5 yielding larger plates limited only by the size of the crucible.  

Electrical transport was measured in a commercial cryostat using standard 4-wire configuration with electrical contact done with Ag paint and Au wires. Magnetization measurements were performed with a commercial SQUID magnetometer using a quartz rod and GE varnish in order to minimize diamagnetic background. Orientation of the single crystals was determined using Laue X-ray diffraction and powder X-ray diffractometer to identify facet orientations. 
To further characterize the crystallographic structure, single-crystal X-ray diffraction using Bruker D8Venture w/ PhotonIII diffractometer was performed on a sample from the Fe:Al=1:10 batch at several temperatures ($T$).

\section{Results}

Single-crystal X-ray diffraction yielded a refined structure consistent with the monoclinic C2/m space group with 6 formula units per unit cell with additional details in table \ref{tab1} and is consistent with previous results \cite{Pop2010Fe4Al13Czhcorlski, Grin1994Kristal}.
We performed temperature-dependent diffraction scans to characterize the crystal structure down to 100~K, finding no evidence of any phase transition nor anomalous structural behavior, with a unit cell volume decreasing monotonically and eventually saturating at approximately 1477 \AA$^3$. Refer to Table 1 for parameters.






\begin{figure*}[]
   \centering
    \includegraphics[scale=0.35]{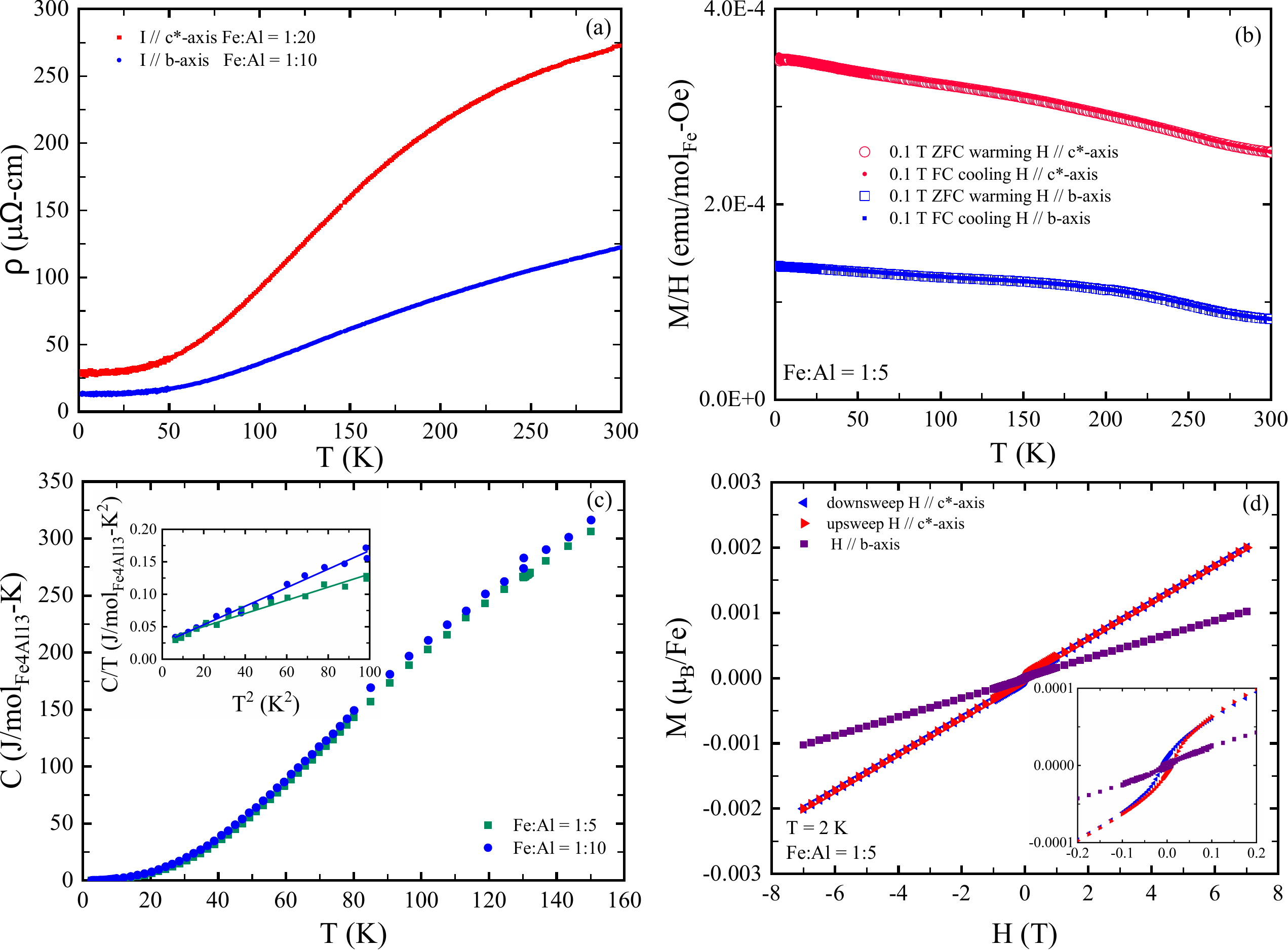}
    \caption{(a) Resistivity ($\rho$) vs temperature ($T$) of Al flux-
grown single crystals of Fe$_4$Al$_{13}$ for current ($I$) along the c* axis (red, Fe:Al = 1:20) and b-axis  (blue, Fe:Al = 1:10).  The overall transport behavior is typical of a metal, although anisotropic owing to low symmetry of crystal structure. (b) Magnetization temperature dependence of a single crystal for growth ratio of Fe:Al = 1:5 taken at 0.1 T along the b-axis and the c*-axis. There is a change in magnitude by changing the field direction, but results are qualitatively similar. Although the magnitude of the magnetization is small, its linear temperature dependence is extremely unusual in that it does not obey Curie-Weiss behavior of localized magnetic moments, nor the constant in $T$ Pauli susceptibility of conduction electrons. The slight asymmetry between ZFC and FC at low $H$ and low $T$ suggest minor contribution from a naturally occurring Fe oxide surface layer. (c) Heat capacity ($C$) vs. $T$ of Al flux-grown Fe$_4$Al$_{13}$ crystals for Fe:Al =1:5 (green) and Fe:Al = 1:10 (blue). The insert shows a low-$T$ inset $C/T$ vs. $T^2$ with fits to the standard electronic and phonon terms. Low Sommerfeld coefficients $\gamma$ indicate a weakly correlated electronic state. There is also a slight difference in the phonon behaviors reflected in the different Debye temperatures between the two samples, which suggest that changes to stoichiometry and growth conditions can alter the phonon spectrum. (d) Magnetization field dependence of single crystals for Fe:Al = 1:5 measured at 2 K. The crystalline anisotropy is minimal between fields along the c*-axis and the b-axis. The inset emphasizes a minor hysteretic contribution from a surface oxide.}
    \label{fig2}
\end{figure*}

The resistivity ($\rho$) vs. temperature ($T$) of \Fal{} crystals for Fe:Al = 1:20 and 1:10 with current ($I$) parallel to the reciprocal lattice $c$*-axis (ie perpendicular to $ab$-plane) and parallel to the $b$-axis, respectively, is shown in Figure~ \ref{fig2}(a). The electrical transport is typical of a paramagnetic metal, with decreasing resistivity on cooling. The magnitude of $\rho(T)$ is much smaller for $I \parallel b$-axis, which may be a reflection of the electronic structure and mobility anisotropy. The magnetoresistance is small for both samples, at most + 10 percent, even at a magnetic field ($H$) of 9 T and 2 K (not shown for brevity), perhaps expected for a system prone to disorder owing to the complicated unit cell of low symmetry and off-stoichiometric tendencies. The Hall effect was measured for the Fe:Al = 1:20 sample ($H$ parallel a-axis) between 2 K and 50 K and shows a single band hole-like response with a temperature independent Hall coefficient of $R_H$ $\sim$ + 1.4E-7 $\Omega$-cm/T, which corresponds to a carrier density of $\sim$ + 4.4E+21 cm$^{-3}$.


The Magnetization ($M$) behavior is shown in Figure~ \ref{fig2}(b) and demonstrates significant departure from expected behavior of both magnetic and non-magnetic metals. A comparison of $M/H$ vs. $T$ for Fe:Al = 1:5  at 0.1 T shows minimal qualitative anisotropy for fields along the b-axis and along the c*-axis. The low field magnetic response is noticeably larger for fields along the c*-axis compared to the b-axis, but there is a hump feature at $\sim$ 225 K for both field directions. At colder $T$ the $M/H$ begins to increase with a linear in $T$, or perhaps slightly sublinear, trend until at the coldest $T$ measured where is a  minor asymmetry between zero field cooled (ZFC) and field cooled (FC) data. This behavior of $M/H$ vs $T$ is extraordinary unusual and does not match anything close to the expected behavior of local moment nor itinerant magnetic response of materials \cite{Mugiraneza2022ComPhys}.

The heat capacity, shown in Figure~ \ref{fig2}(c), is again typical of a weakly correlated metal, with a small $T$-linear component at low temperatures consistent with a nearly-free electron mass uncomplicated by correlations or other sources of entropy enhancement. 
A fit with the standard conduction electron Sommerfeld ($\gamma T$) and phonon Debye ($BT^3$) terms below 10 K is shown in the inset with the 1:10 sample having slightly smaller $\gamma$ than the 1:5 sample, consistent with a minor change in the density of states at the Fermi energy. The values of $\gamma$ for the Fe:Al = 1:5 and 1:10 samples are 7.5 $\pm$ 0.6 mJ/mol-K$^2$ and 6.1 $\pm$ 0.6 mJ/mol-K$^2$, respectively. In contrast, the 1:5 sample has a larger Debye temperature $\theta_D$, which suggest changes in stoichiometry also alters the phonon spectrum.  The values of $\theta_D$ for the Fe:Al = 1:5 and 1:10 samples are 319 $\pm$ 5 K and 284 $\pm$ 3 K, respectively. The stoichiometry also alters the higher temperature behavior with the 1:10 having more heat capacity than the 1:5 sample, even up to 150 K. There are no indications of magnetic contributions to heat capacity.

The $M$ vs $H$ behavior at 2 K in Figure~ \ref{fig2}(d) is again suggestive of paramagnetism of dilute Fe moments. The anisotropy is consistent with the results in Figure \ref{fig2}(b) with a larger response with $H$ along the c*-axis compared to b-axis. The initial low $H$ behavior is due to a small amount of ferromagnetic or superparamagnetic naturally occurring Fe oxide on the surface of the crystal.

\begin{figure}[]
    \centering
    \includegraphics[scale=0.3]{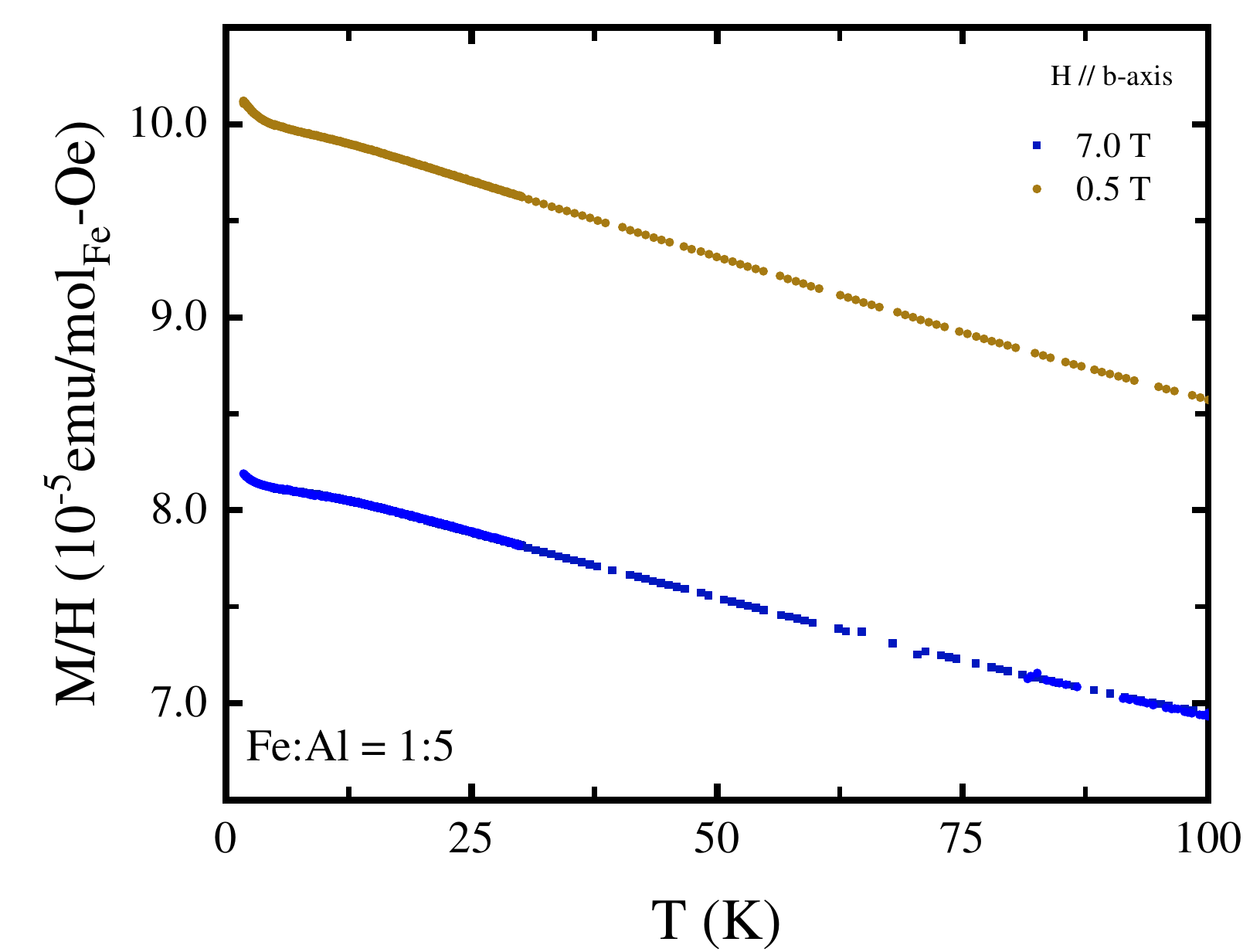}
    \caption{Low Temperature view of $M/H$ vs $T$ for stronger magnetic fields parallel to the b-axis. The linear in $T$ behavior persists to the highest fields measured.}
    \label{fig3}
\end{figure}

The $M/H$ vs $T$ behavior at larger field strengths exhibits minimal qualitative change as shown in Figure \ref{fig3}. The linear in $T$ behavior remains prominent, which suggests that the Zeeman energy scale is the dominant contribution to the thermodynamics. All the above experimental results are consistent with a weakly interacting system, except the $M/H$ vs $T$ scaling and motivates a closer look at the band structure. 
\section{Density Functional Theory Calculations}

\begin{figure}[H]
    \centering
    \includegraphics[scale=0.46]{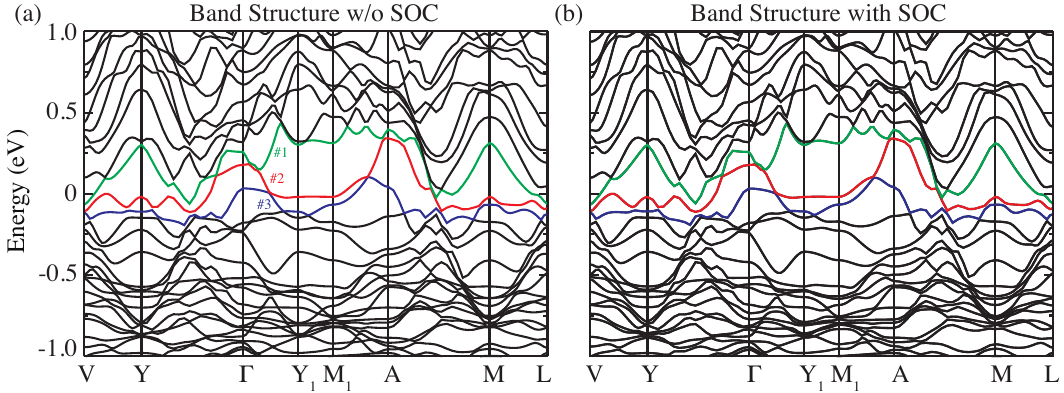}
    \caption{Band structure calculations (a) without and (b) with spin-orbit coupling. The three bands $\#1\sim3$ are indicated with green, red and blue lines. We note the rather flat half filled band (red) hugging the Fermi level.}
    \label{figDFT1}
\end{figure}

Although previous calculations assume ferromagnetic ground state \cite{Israr2018doi:10.1142/S0217979218502016}, we do not observe experimental signatures of a bulk ferromagnetic ground state in the aforementioned results. This motivated another attempt to perform density functional theory (DFT) calculations \cite{vasp1,vasp2} with the results shown in Fig. \ref{figDFT1}. The results without (a) and with (b) spin-orbit coupling (SOC) show minimal apparent difference owing to the light elements involved in the system. As expected, Fe$_{4}$Al$_{13}$ is a metal with an odd number of electrons per primitive cell. There are three bands at the Fermi level that we refer them as band \# 1, \# 2, and \# 3. The band \# 2 that contributes the most at the Fermi level is half-filled. 
We further calculated the band representations at the high-symmetry points in the Brillouin zone and identified the corresponding band topologies for each band \cite{tqc}. In the absence of SOC, none of the three bands (\# 1$\sim$ 3) is isolated because their associated band representations do not satisfy all the compatibility relations of this group. Consequently, symmetry-protected nodal points (or lines) must exist between these bands and those above and below them. Detailed analysis reveals that a mirror-symmetry protected nodal line is present between bands \# 2 and \# 3 (as well as between bands \# 2 and \# 1) on the G-A-M-Y plane.
With SOC included, the nodal lines become gapped and all three bands are isolated. For each gap, we consider all bands below it as occupied and calculate the corresponding topological indices. The results indicate that the gap between bands \# 1 and \# 2 is topologically nontrivial with indices $(z_{2,1}, z_{2,2}, z_{2,3}, z_4) = (0003)$, while the gap between bands \# 2 and \# 3 is also topologically nontrivial with indices $(z_{2,1}, z_{2,2}, z_{2,3}, z_4) = (1101)$. Both of the above indices indicate a strong topological insulator phase.
However, the presence of a multitude of bands (due to the large number of atoms per unit cell and the low symmetry group) gives rise to a complicated "spaghetti-like" band structure, with strong metallic character. As such the topological indices should be interpreted as indices of the bands rather than those of the Fermi level which again, is metallic.  We point out that the red-band which stoichiometrically is half-filled, is much flatter than many of its neighbors, with bandwidth less than 100 meV for most momenta. Half filled flat bands have strong tendency to magnetize, shown both experimentally, and theoretically. Hence experimental observation of lack of magnetism in this compound is surprising.

\begin{figure}[H]
    \centering
    \includegraphics[scale=0.4]{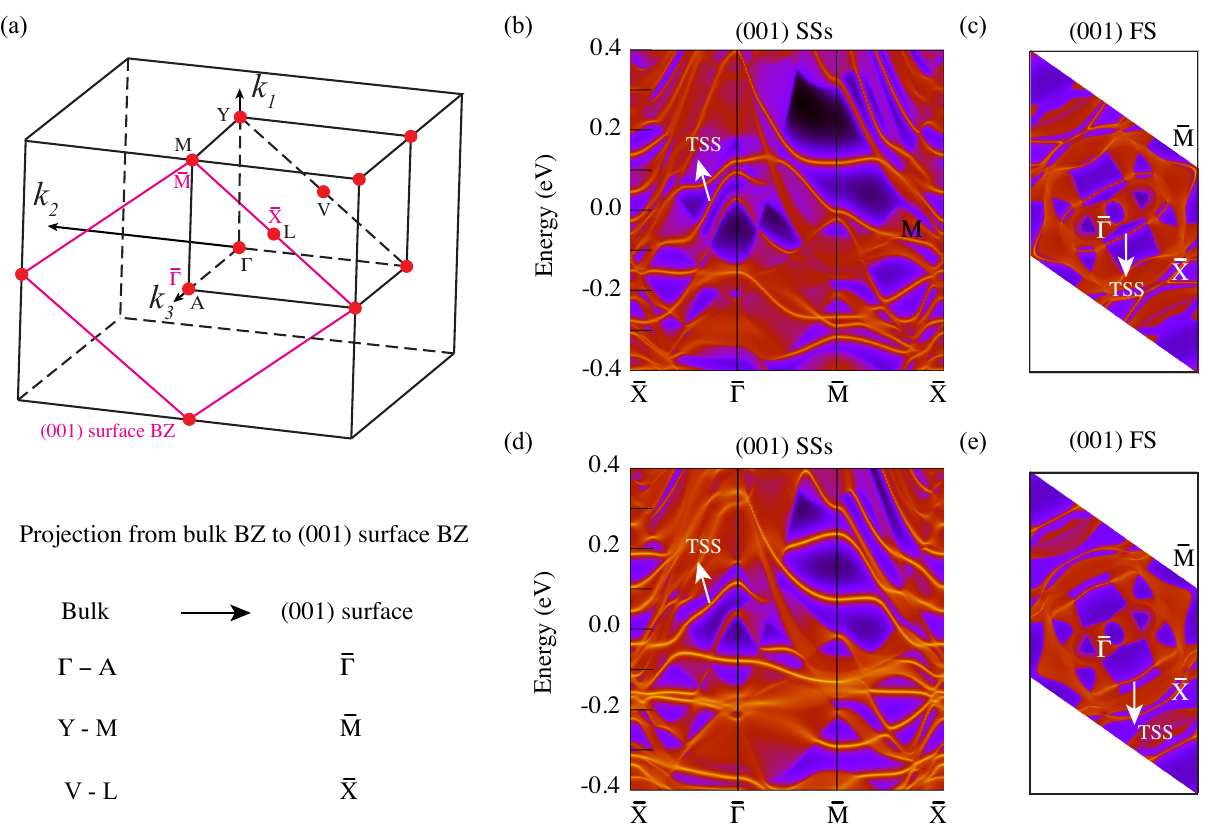}
    \caption{Surface states (SSs) calculations. (a) The bulk Brillouin zone (BZ) with high-symmetry points labeled is shown, with the (001) surface BZ highlighted in magenta. Below, an illustration demonstrates how the high-symmetry points are mapped from the bulk to the (001) surface. The three reciprocal lattice vectors are defined as $(\mathbf{k_1,k_2,k_3})=\frac{2\pi}{\Omega}(\mathbf{b\times c, c\times a, a\times b})$, where $\mathbf{(a,b,c)}$ represent the three lattice vectors shown in Fig. \ref{fig1} and $\Omega$ denotes the volume of the unit cell. Band dispersion (b) and Fermi surface (FS) (c) on the (001) surface with Al termination are displayed, with topological surface states (TSS) along the $\Bar{X}-\Bar{\Gamma}$ path clearly indicated in both plots. Panels (d) and (e) show similar results for the Fe-terminated (001) surface.}
    \label{figDFT3}
\end{figure}

We computed the surface states along the (001) direction using the Green’s function method \cite{wanniertools}. Fig. \ref{figDFT3}(a) displays the first Brillouin zone with labeled high-symmetry points, along with an illustration of how these points map from the bulk to the (001) surface Brillouin zone. Figs. \ref{figDFT3}(b) and (c) present the dispersion of the surface states along high-symmetry points and the Fermi surface on the (001) surface (with Al termination), respectively, clearly distinguishing the topological surface states from the bulk states along $\Bar{X} -\Bar{\Gamma}$. Similar results were obtained for the Fe-terminated surface in Figs. \ref{figDFT3}(d) and (e).


\section{Discussion}

This work demonstrated the utility of \Fal{} crystal growth for educational purposes with an Al flux growth method, revealing a surprising lack of magnetism and an unusual magnetic susceptibility behavior. The electrical transport and heat capacity suggest a weakly correlated non-magnetically ordered metal. The magnetization, however, demonstrates anomalous temperature dependence that is inconsistent with local moment Curie-Weiss behavior and itinerant Pauli paramagnetism. A comparison of our results with Czochralski-grown \Fal{} crystals \cite{Pop2010Fe4Al13Czhcorlski, Gille2008CrystalResearch} shows similar transport and heat capacity, although their magnetization results are at first glance more suggestive of Curie-Weiss-like behavior. This suggests that \Fal{} crystals by slow cooling Al flux create an interesting collection of dilute Fe moments that breaks conventional weakly correlated paramagnetic behavior. Although \Fal{} is crystalline, it is very close to being a decahedral quasicrystal and is described as a quasicrystal approximant \cite{Liu2024Metals}. As we observe no indication of single ion Kondo behavior \cite{HEW.93}, the local Fe impurity moment coupling with the conduction electrons must be very weak. As the band structure is suggested to be complicated due to the Hall and thermopower response \cite{Pop2010Fe4Al13Czhcorlski}, it can be expected that the Ruderman–Kittel–Kasuya–Yosida (RKKY) \cite{Xin2021PhysRevB.104.144421} conduction electron mediated interaction among the dilute Fe moments is non-trivial.  
In fact, it has already been confirmed that tuning the RKKY interaction in quasicrystal approximant Au-Ga-Tb systems can change the ground state from ferromagnetic-like to antiferromagnetic-like \cite{LABIB2024101321}. It is conceivable that  our \Fal{} is a more dilute version of this in which we have random antiferromagnetic and ferromagnetic RKKY mediated interactions among Fe impurities, which leads to frustration and breakdown in Curie-Weiss behavior. 
It suggests future examination of dilute paramagnetic behavior in metallic quasicrystals and quasicrystal approximants could be worthwhile in order to test for breakdowns in the very fundamental Curie-Weiss law, although we note the even J. H Van Vleck himself pointed out the limits of the Curie-Weiss law \cite{VANVLECK1973177}.

\begin{acknowledgments}
The authors acknowledge generous continued support from funders of the FQM Winter School, in particular the Gordon and Betty Moore Foundation. We also thank the students of the January 2024 Fundamentals of Quantum Materials Winter School who synthesized a significant quantity of \Fal{} crystals as part of their hands-on module training.

\end{acknowledgments} 
%


\end{document}